\documentclass[preprint,aps,pre]{revtex4-1}
\usepackage{graphicx}
\usepackage{fancybox}
\usepackage{amsmath,amssymb,bm,amsthm}
\def\beq{\begin{equation}}
\def\eeq{\end{equation}}

\def\<{\langle}
\def\>{\rangle}

\begin{document}

\title{Complete catalog of ground-state diagrams for the general three-state 
lattice-gas model with nearest-neighbor interactions on a square lattice}

\author{Daniel Silva$^1$ and Per Arne Rikvold$^{1,2}$}
\affiliation{$^1$ Department of Physics, Florida State University, 
Tallahassee, FL 32306-4350,USA\\
$^2$ PoreLab, NJORD Centre, Department of Physics, University of Oslo, P.O.\ Box 1048 Blindern, 
0316 Oslo, Norway
}

\begin{abstract}
The ground states of the general three-state lattice-gas (equivalently, $S=1$ Ising) model 
with nearest-neighbor interactions on a square lattice are explored in the full, five-dimensional 
parameter space of three interaction constants and two generalized chemical potentials or fields. 
The resulting, complete catalog of fifteen topologically different ground-state diagrams 
(zero-temperature phase diagrams) is 
discussed in both lattice-gas and Ising-spin language. The results extend those of 
a recent study in a reduced parameter space [V.\ F.\ Fefelov, {\it et al.\/}, 
Phys.\ Chem.\ Chem.\ Phys., 2018, {\bf 20}, 10359--10368], which identified 
six different ground-state diagrams. 
\end{abstract}

%\pacs{75.30.Wx,64.60.My,64.60.Kw,75.60.-d}

\maketitle

%*******************************************************************************
%\begin{document}

\section{Introduction}
\label{sec:I}

Many physical, chemical, or social systems can be described as a lattice or a more general 
network, whose sites carry a discrete variable with a finite number of possible states. The simplest 
example is the $S=1/2$ Ising model \cite{ISING25,ONSA44,IBAR16}, in which each site can take one of two values. 
Common interpretations in physics and chemistry include opposite directions of a magnetic 
or electric dipole moment or ``spin" ($\sigma \in \{-1,+1\}$), 
or the absence or presence of an atom or a molecule ($c \in \{0,1\}$). 
In the social sciences, the states might, e.g., represent opinions, population groups, 
or languages \cite{STAU08}. 

In the present paper we consider the more complex case of three possible states, 
known as the $S=1$ Ising model ($\sigma \in \{-1,0,+1\}$) or, equivalently, the three-state 
lattice-gas (LG) model with states A, B, and vacancy (0). The spin representation has been 
used to describe magnetic or dielectric systems with local triplet states \cite{CAPE66} or 
%(with less success) 
the $\lambda$ transition in He$^3$-He$^4$ mixtures \cite{BLUME71}. 
In its LG form, the model has been used extensively to study two-component adsorption at 
solid-gas \cite{FEFE18} and solid-liquid \cite{RIKV88B,COLL89,RIKV96} interfaces, 
as well as spatially complex mixtures 
of protons, neutrons, and voids (``nuclear pasta") thought to exist in the inner crust of 
neutron stars \cite{HASN13}. 

Models using such discrete-state representations are typically defined by an effective  
Hamiltonian including  ``interaction constants" representing the tendency of 
state variables on neighboring sites to take equal or different values, and external 
``chemical potentials"  or ``fields" that determine the individual energies of the state variables. 
While the two-state model with nearest-neighbor interactions contains one interaction constant 
and one field, the general three-state model contains three interaction constants and 
two fields. The tendency to seek an ordered minimum-energy configuration, 
or {\it ground state\/}, is counterbalanced by a ``temperature" parameter encouraging disorder. 
Due to this competition between order and disorder, complex 
phenomena, including continuous or discontinuous  phase transitions, occur at nonzero temperatures. However, aspects of the ground-state configurations are typically observable (at least as local fluctuations), even at quite high temperatures. A study of the 
{\it ground-state diagram\/} (zero-temperature phase diagram) formed by the lines or planes in a parameter space of interaction 
constants and fields that separate different ground states is therefore a natural 
starting point for detailed studies of 
equilibrium or nonequilibrium phenomena at nonzero temperatures. 
The ground-state diagram can be likened to the foundation, on which the finite-temperature 
``building" representing the full phase diagram is supported. 

For the two-dimensional triangular lattice, complete analysis of the topologically different 
ground-state diagrams produced by different parameter values in the three-state LG model with 
nearest-neighbor interactions and the equivalent $S=1$ Ising  
model were presented in Refs.\ \cite{RIKV88B} and \cite{COLL88}, respectively. 
For the simpler, two-dimensional square lattice, ground-state diagrams for a related LG model with nearest- and next-nearest neighbor repulsive interactions were obtained by 
Huckaby and Kowalski \cite{HUCK84}. 
Very recently, Fefelov {\it et al.\/} presented the six
possible phase diagrams for the three-state LG model in the special case of vanishing 
interactions between particles of opposite kinds, representing an additive gas mixture \cite{FEFE18}. 
However, to the best of our knowledge, 
no full enumeration of topologically different ground-state diagrams for the square lattice 
in the general case of 
three non-vanishing interaction constants has previously been presented. 
The purpose of the present paper is to provide such a complete description in both LG and Ising-spin 
language for this important lattice, whose physical realizations include the (100) planes of the 
three-dimensional body-centered and face-centered cubic crystal lattices.  

The rest of this paper is organized as follows. 
The model Hamiltonian is presented in Sec.\ \ref{sec:H}, with the spin formulation in 
Subsec.\ \ref{sec:MAG} and the LG formulation in Subsec.\ \ref{sec:LG}, where the 
relations between the model parameters in the two representations are also given. 
Section \ref{gscalc} consists of three subsections. 
The six possible ground-state configurations 
are reviewed in Subsec.\ \ref{gs}, and phase diagrams in 
six different asymptotic strong-field limits are described in Subsec.\ \ref{asymp}. 
Our main results, the complete enumeration and description of the fifteen topologically different 
ground-state diagrams in the intermediate and weak field limits where the asymptotic regions meet, 
are given in Subsec.\ \ref{topol}. 
A short summary and our conclusions are given in Sec.\ \ref{sec:CONC}.

\section{Model}
\label{sec:H}
\subsection{Ising-spin formulation}
\label{sec:MAG}
The most general form for the  $S = 1$ Ising Hamiltonian with nearest-neighbor 
pairwise interactions takes the form,
\begin{eqnarray} 
\label{eq:BEG}
\mathcal{H}_{\text{$S=1$}} 
&=& -J \sum_{\langle i,j \rangle} p_i p_j - K \sum_{\langle i,j \rangle} q_i q_j - 
L \sum_{\langle i,j \rangle} (q_i p_j + p_i q_j) \nonumber\\
&& + D \sum_i q_i - H \sum_{i} p_i \;.
\end{eqnarray}
Here, $\sum_{\langle i,j \rangle} $ denotes summation over all nearest-neighbor pairs, where $  p_i \in  \{ -1,0,1 \}  $ and  $q_i = p_i^2  $. The parameters $J$ and $H$ are analogous to the 
single interaction constant and the external field in the spin 1/2 Ising model, respectively. 
[See Eq.\ (\ref{eq:ising}) below.] 
The interaction constant $J$ $ > $ 0  and $J < 0  $ correspond to the 
uniform ferromagnetic (FM) case and the checkerboard antiferromagnetic (AFM) case, respectively, and the field $H$ distinguishes between positive and negative $p_i$. The ``crystal field" $D$ distinguishes between zero and nonzero $p_i $, with $D <  0$ denoting preference 
for $ q_i = 1 $. $K <  0$ denotes preference for bonds with at least one zero spin, while $K >  0$ denotes a preference for nearest neighbors being nonzero, irrespective of sign.  $L > 0$ 
corresponds to a preference for ferromagnetic ordering with $p_i = +1 $. 
This most general $S=1$ Ising model is invariant under the transformation 
$L \to -L, H \to - H, p_i \to -p_i$. Its ground states 
and ground-state diagrams were studied on a triangular lattice in 
Ref.\ \cite{COLL88}. 

Setting $L=0$ one gets the Blume-Emery-Griffiths (BEG) model \cite{BLUME71}, while
setting $K = L=0$ leads to the Blume-Capel (BC) model \cite{CAPE66}. 
Additionally setting $D=0$ and limiting the spins to up or down, $\sigma_i \in \{ -1,+1 \} $, 
one obtains 
the $S= 1/2 $ Ising model, known as the ``hobby horse" of magnetic systems \cite{IBAR16},
\begin{equation} \label{eq:ising}
\mathcal {H}_{\text{Ising}} 
= -J \sum_{\langle i,j \rangle} \sigma_i \sigma_j - H \sum_{i} \sigma_i \;.
\end{equation}
%In the following we present an exhaustive enumeration of the possible ground states on a square lattice.

 \subsection{Lattice-gas formulation}
\label{sec:LG}
The $S = 1$ Ising model can be mapped to a three-state (A, B, and vacancy 0) LG model. It can represent two gases with molecules of types A and B or two solutes and a solvent. Interactions between molecules of types A and B are denoted $\phi_{AB}$, while interactions between molecules of same type are denoted $\phi_{AA}$ and $\phi_{BB}$. The Ising Hamiltonian can be transformed to a lattice-gas Hamiltonian by introducing the local concentration variables, 
\begin{equation} 
c_i^A = \frac{1}{2}(q_i +p_i) 
\label{eq:cA}
\end{equation} 
and
\begin{equation}
c_i^B = \frac{1}{2}(q_i - p_i)  \;,
\end{equation} 
and defining the interaction energies $ \phi_{AA} ,\phi_{BB},$ and $ \phi_{AB} $ 
and the chemical potentials $ \mu_A $ and $\mu_B $ ,  as \cite{COLL88} 
\begin{eqnarray}
\label{eq:PHI}
\phi_{AA} &=& J +K + 2L \nonumber\\
\phi_{BB} &=& J +K - 2L \nonumber\\
\phi_{AB} &=& K-J \\
\mu_A &=& H -D \nonumber\\
\mu_B &=& -H-D \;. \nonumber
\end{eqnarray} 
From these definitions we get the grand-canonical LG Hamiltonian,
\begin{eqnarray}
\mathcal{H}_{\rm LG} &=& -\phi_{AA} \sum_{\langle  i,j \rangle}c_i^A c_j^A - \phi_{BB} \sum_{\langle  i,j \rangle}c_i^B c_j^B 
          - \phi_{AB} \sum_{  \langle i,j  \rangle}(c_i^A c_j^B + c_i^B c_j^A) \nonumber\\
 && -\mu_A \sum_{i} c_i^A - \mu_B \sum_{i} c_i^B  \;.
\label{eq:HLG}
\end{eqnarray} 
The special case of $\phi_{AB} = 0$, or equivalently $K=J$, was recently studied 
in Ref.\ \cite{FEFE18}.

The mapping defined by Eqs.\ (\ref{eq:cA}) -- (\ref{eq:PHI}) is trivially inverted to yield 
\begin{eqnarray} 
\label{eq:pq}
p_i &=&c_i^A - c_i^B \nonumber\\
q_i &=& c_i^A + c_i^B  
\end{eqnarray} 
and
\begin{eqnarray}
J &=& \frac{1}{4} \left(\phi_{AA} -2 \phi_{AB} + \phi_{BB} \right) \nonumber\\
K &=& \frac{1}{4} \left(\phi_{AA} +2 \phi_{AB} + \phi_{BB} \right) \nonumber\\
L &=& \frac{1}{4} \left(\phi_{AA} - \phi_{BB} \right) \\
D &=& - \frac{1}{2} \left( \mu_A + \mu_B \right) \nonumber\\
H &=&  \frac{1}{2} \left( \mu_A - \mu_B \right) \;. \nonumber
\end{eqnarray} 

\section{Ground-state calculation} 
\label{gscalc}
\subsection{Ground states}
\label{gs}

The density conjugate to the chemical potential $\mu_X$ is 
the coverage $\theta_X = N^{-1} \sum_i c_i^X $, where $N$ is the 
total number of lattice sites. From Eq.\ (\ref{eq:pq}) we note that $p_i = +1$ means $c_i^A =1$ 
and $c_i^B = 0$, and opposite for $p_i = -1$. 
The macroscopic densities conjugate to the fields $H$ and $-D$ are the magnetization 
$P = N^{-1} \sum_i p_i$ and the quadrupole moment $Q = N^{-1} \sum_i q_i$, respectively. 
Specific phases are identified by their corresponding 
values of $P$ and $Q$ by $(X  \times Y)_P^Q $. Here, $X$ and $Y$ denote the periodicities 
in the two lattice directions, a notation common in surface science. 
The energy per lattice site of a particular phase is found by evaluating the Hamiltonian for the corresponding configuration,
\begin{equation}
E_{(X  \times Y)_P^Q} = \cfrac{\mathcal{H}_{(X  \times Y)_P^Q}}{N} \;.
\end{equation}
The ground state is the phase with the minimum energy. 
It is a function of the parameters $J,K,L,D,$ and $H$ 
(or $\phi_{AA}, \phi_{BB}, \phi_{AB},\mu_A, \text{and } \mu_B)$, 
\begin{equation} 
E_{\rm gs} (J,K,L,D,H ) = \text{ min}\{E_{(X  \times Y)_P^Q}  \} \;. 
\end{equation}

\begin{figure}
\[
 \begin{matrix}
   \includegraphics[width=2cm,height=3cm,keepaspectratio]{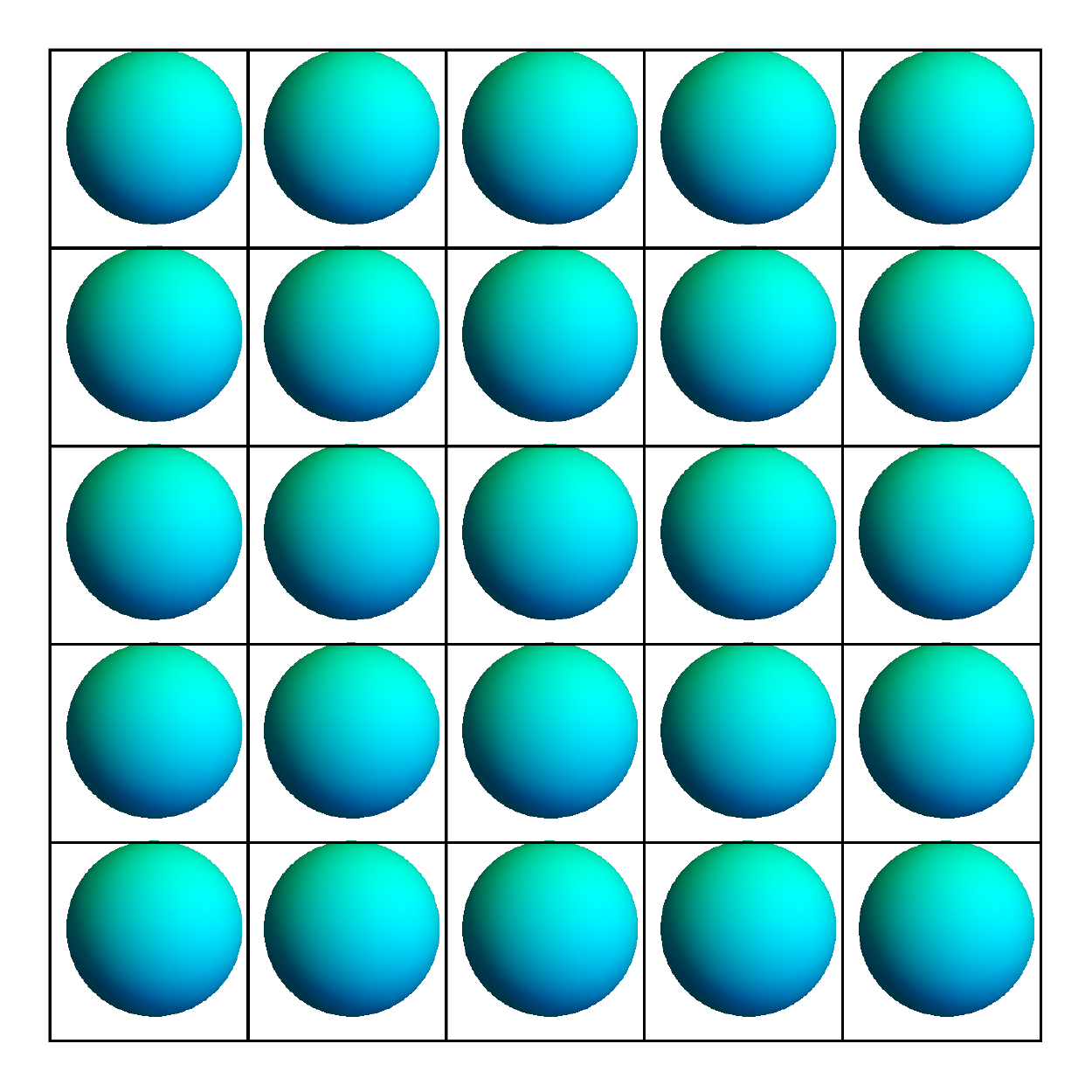}&
     \includegraphics[width=2cm,height=3cm,keepaspectratio]{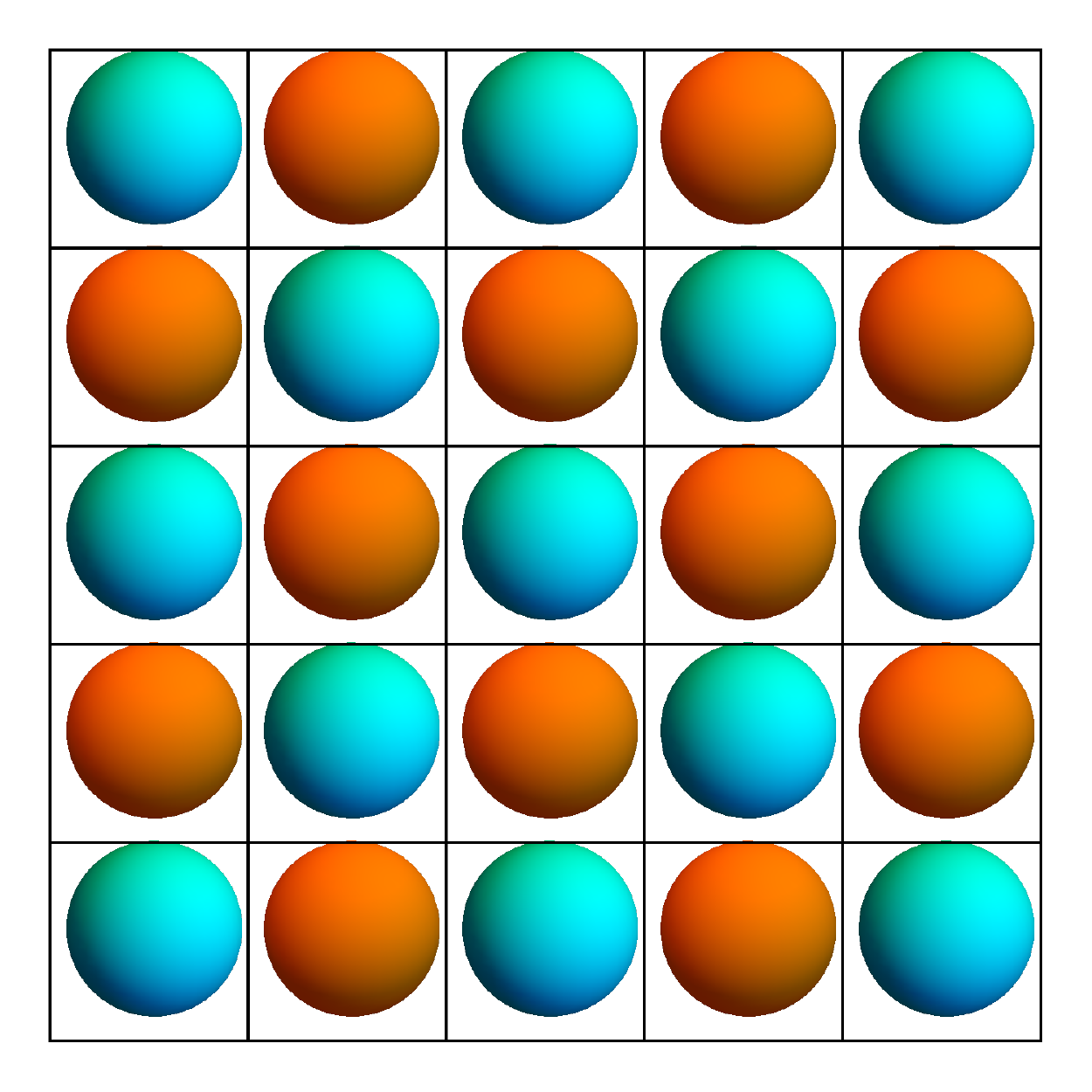} \\
       (1\times1)_1^1 &    (\sqrt{2}\times\sqrt{2})_{0}^1   \\
   \includegraphics[width=2cm,height=3cm,keepaspectratio]{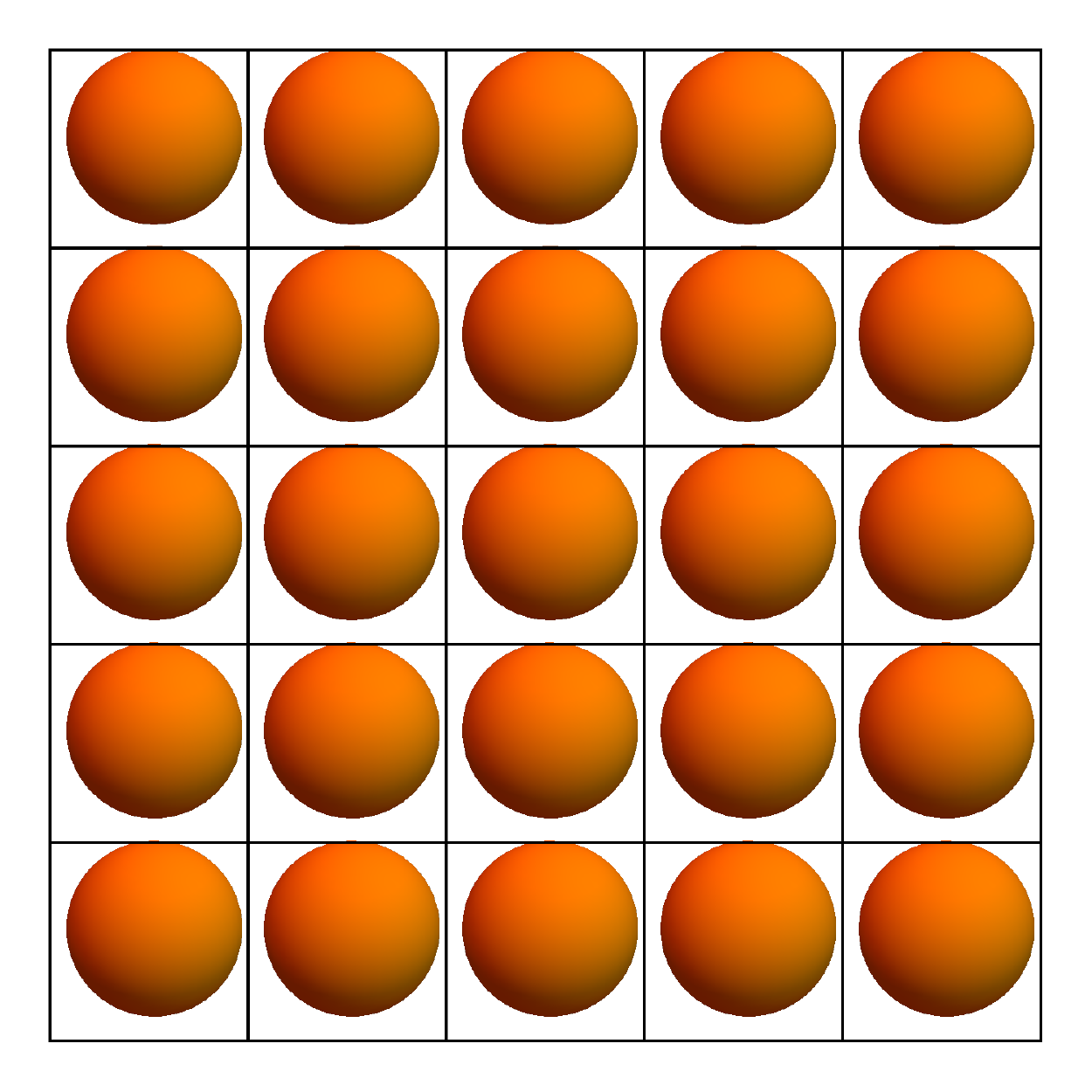}&
    \includegraphics[width=2cm,height=3cm,keepaspectratio]{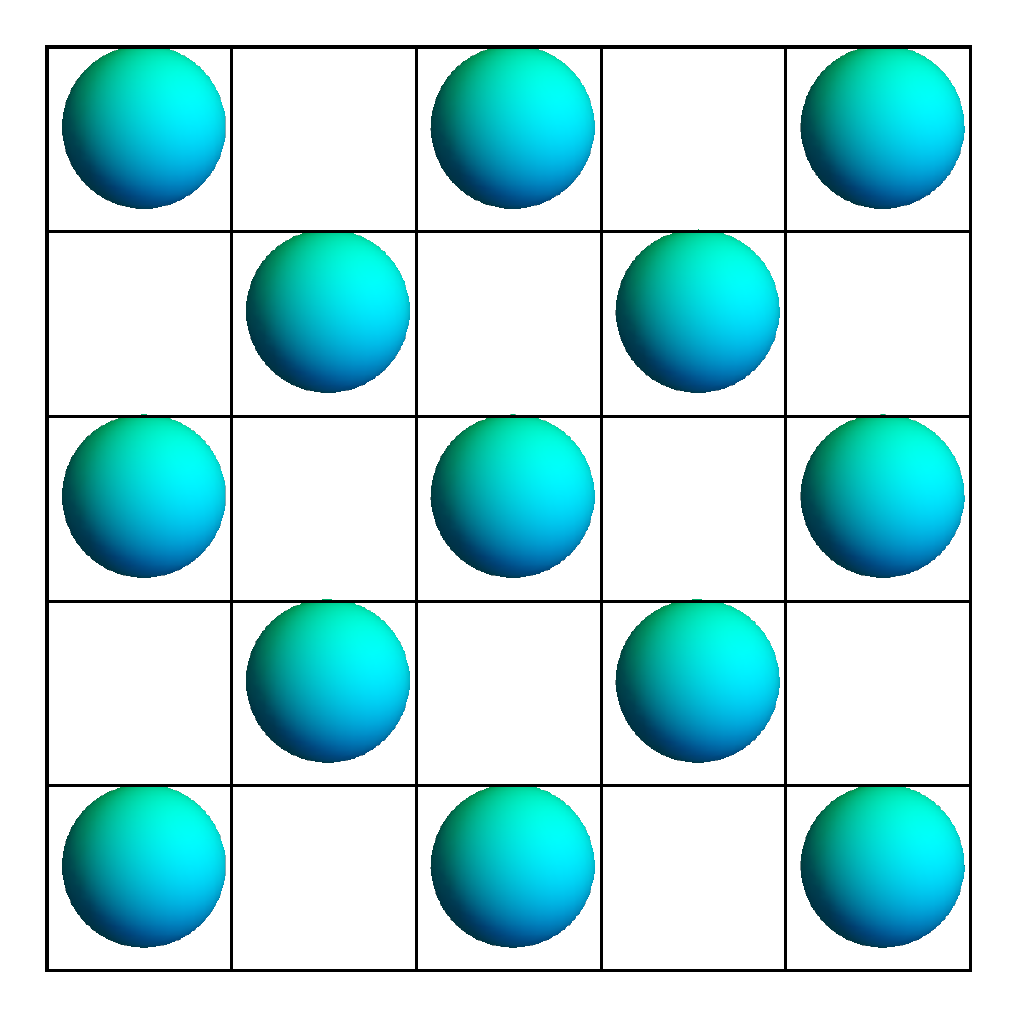}\\ 
    (1\times1)_{-1}^1 & (\sqrt{2}\times\sqrt{2})_{1/2}^{1/2}  \\
     \includegraphics[width=2cm,height=3cm,keepaspectratio]{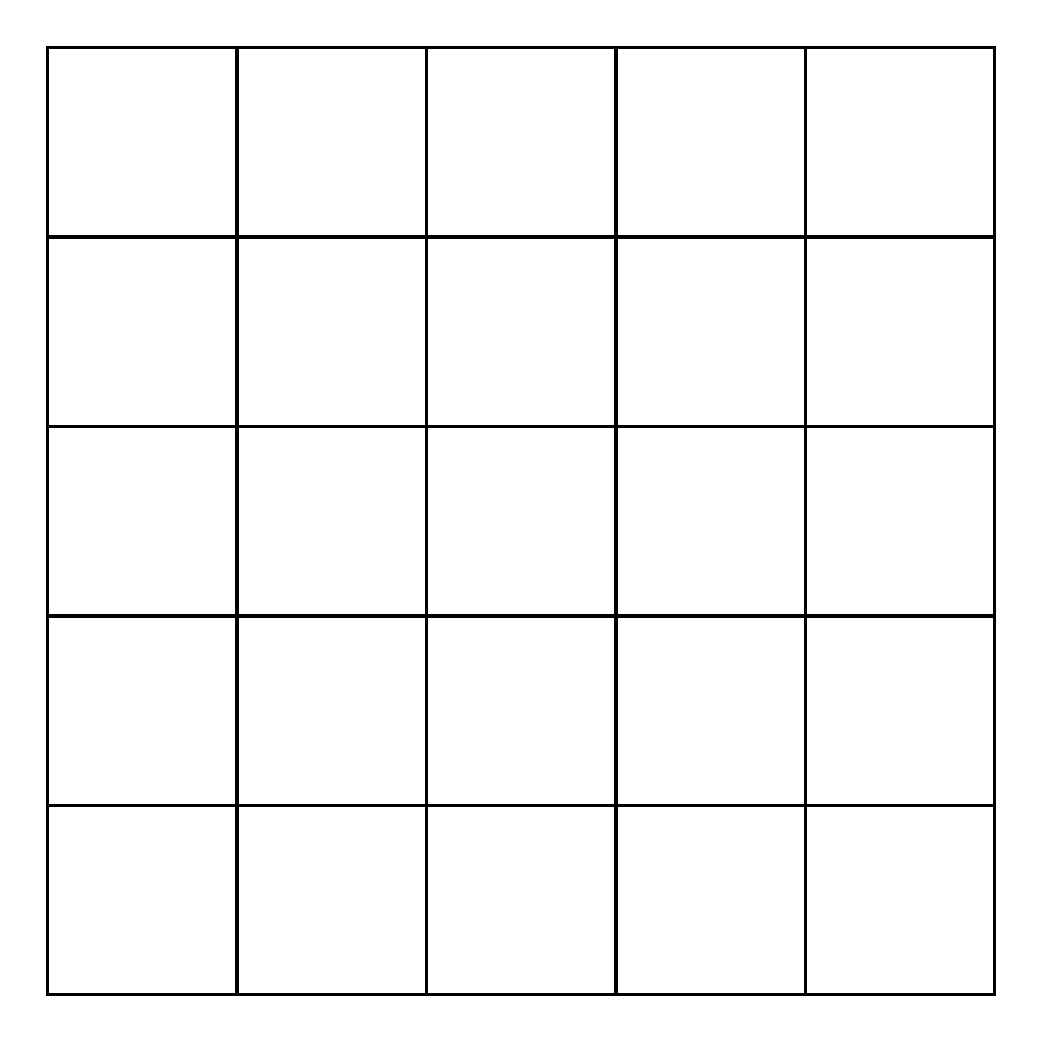} &
     \includegraphics[width=2cm,height=3cm,keepaspectratio]{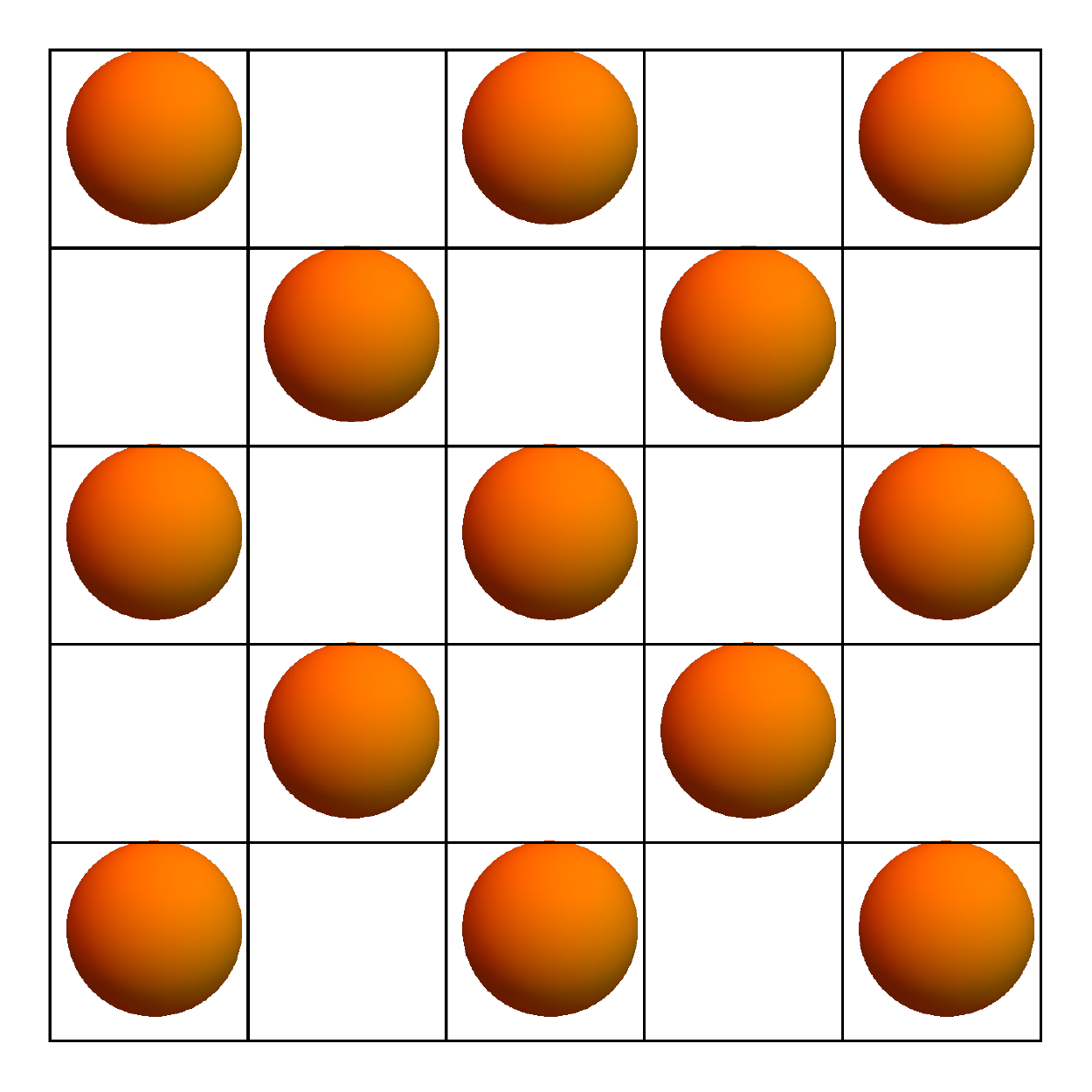} \\

       (1\times1)_{0}^0 &     (\sqrt{2}\times\sqrt{2})_{-1/2}^{1/2}  \\
 \end{matrix}
\]
\caption{The six possible ground-state configurations 
for the model defined in Eqs.\ (\ref{eq:BEG}) and  (\ref{eq:HLG}). The disordered (uniform) 
phases are $ (1\times1)_1^1$ (all A),  $(1\times1)_{-1}^1$ (all B), and $(1\times1)_{0}^0$ 
(empty lattice). 
The ordered phases are $ (\sqrt{2}\times\sqrt{2})_{0}^1 $ (A and B), 
$ (\sqrt{2}\times\sqrt{2})_{1/2}^{1/2} $ (A and 0), 
and  $(\sqrt{2}\times\sqrt{2})_{-1/2}^{1/2}$ (B and 0). 
}
\label{fig:pgs}
\end{figure}
Candidate ordered phases are chosen among those that can be reached from the disordered 
$(1 \times 1)$ phases by continuous phase transitions, as determined by group-theoretical 
arguments \cite{DOMA78,SCHI81}. In the absence of interactions beyond nearest neighbors, the 
only ordered phases possible have the $(\sqrt{2} \times \sqrt{2})$ symmetry \cite{SCHI81}, 
which divides the lattice into two interpenetrating sublattices. 
In spin language, these checkerboard phases are commonly called {\it antiferromagnetic\/} (AFM). 
For different values of the five parameters,
six different phases can be formed. These are shown in Fig.\ \ref{fig:pgs}. There are three 
uniform (disordered) phases, $(1\times1)_1^1$ (all A), $(1\times1)_{-1}^{1}$ (all B), and 
the empty lattice $(1\times1)_0^0$ (all 0), and three ordered checkerboard (AFM) phases, 
$(\sqrt{2}\times\sqrt{2})_{0}^{1}$ (A and B), 
$(\sqrt{2}\times\sqrt{2})_{1/2}^{1/2}$ (A and 0), and 
$(\sqrt{2}\times\sqrt{2})_{-1/2}^{1/2}$ (B and 0).
These ground states, along with their corresponding values of $Q,P, \theta_A, \theta_B$ and their energies per lattice site, are shown in Table \ref{table:gs}, using both Ising and LG notation.

\begin{center}  
\begin{table}
\caption{The ground states and their properties, in both Ising and LG language.
The ground state $(\sqrt{2}\times\sqrt{2})_{0}^1 $ exists only in the AFM case ($J < 0$).}
\label{table:gs}
  \begin{tabular}{ l  c  c c c c c}
    \hline
    \hline
    State & Config. & $Q$ & $P$ & $\theta_A$ & $\theta_B$  & Energy per site \\ \hline
    $(1\times1)_1^1 $& $\begin{matrix} AA \\ AA \end{matrix}$  & 1 & 1 & 1 & 0 & $\begin{matrix} -2J-2K-4L+D-H \\ =-2\phi_{AA} - \mu_A \end{matrix}$ \\ \hline
    $(1\times1)_{-1}^1 $ & $\begin{matrix} BB \\ BB \end{matrix}$ & 1 & $-1$ & 0 & 1 & $\begin{matrix} -2J-2K+4L+D+H \\ = -2\phi_{BB} - \mu_B \end{matrix}$\\ \hline
    $(1\times1)_{0}^0 $ & $\begin{matrix} 00 \\ 00 \end{matrix}$ & 0 & 0 & 0 & 0 & 0\\ \hline
     $(\sqrt{2}\times\sqrt{2})_{0}^1 $ & $\begin{matrix} AB \\ BA \end{matrix}$ & 1 & 0 & 1/2 & 1/2 & $\begin{matrix} 2J-2K+D \\= -2\phi_{AB} - 
     \cfrac{1}{2}(\mu_A +\mu_B) \end{matrix}$\\ \hline     
      $(\sqrt{2}\times\sqrt{2})_{1/2}^{1/2} $ & $\begin{matrix} A0 \\ 0A \end{matrix}$ & 1/2 & 1/2 & 1/2 & 0 & $\begin{matrix}  \cfrac{1}{2}(D-H)= -\cfrac{1}{2}\mu_{A}  
     \end{matrix}$\\ \hline
      $(\sqrt{2}\times\sqrt{2})_{-1/2}^{1/2} $ & $\begin{matrix} B0 \\ 0B \end{matrix}$ & 1/2 & $-1/2$ & 0 & 1/2 & $\begin{matrix}  \cfrac{1}{2}(D+H)= -\cfrac{1}{2}\mu_{B}  
     \end{matrix}$\\
    \hline
    \hline
  \end{tabular}
\end{table}
\end{center}

\subsection{Asymptotic results}
\label{asymp}

The phase boundaries in the five-dimensional parameter space are found by 
pair-wise equating the energy per site of different phases.  
The ground state is uniform in three asymptotic strong-field limits. 
As $H \to +\infty$, the ground state is  $(1\times1)_1^1$, 
while for $H \to -\infty$ it is $(1\times1)_{-1}^1 $. Similarly, for $D \to +\infty$, 
empty sites become energetically favorable, and the ground state is $(1\times1)_{0}^0 $. 

In three other asymptotic directions, the Hamiltonian reduces to the $S=1/2$ Ising model, Eq.\ (\ref{eq:ising}), but with effective parameters, 
$\hat{J}$ and $\hat{H}$, which are functions of the parameters of the full Hamiltonian, 
\begin{equation}
\mathcal{\hat{H}_{I}} = -\hat{J}  \sum_{\langle  i,j \rangle} \sigma_i \sigma_j 
- \hat{H} \sum_i \sigma_i \;.
\label{eq:iseff} 
\end{equation} 

As $D \to - \infty$, $ p_i = q_i = 0$ becomes energetically unfavorable, 
and the Hamiltonian reduces to Eq.\ (\ref{eq:iseff}) with
\begin{equation}
\begin{split}
\sigma_i &= p_i = \pm1 \\
\hat{J}&=J \\
\hat{H}& = H + 4 L \;.
\end{split}
\end{equation}
This is just the $S=1/2$ Ising model with the original interaction constant $J$
and the field shifted by $- 4L$. It is FM or AFM, depending on whether $J$ is positive or negative, 
respectively.  

As $H \to + \infty$ and $D \to + \infty$ $ ( \mu_B \to -\infty)$ it is energetically unfavorable to have $p_i = -1$ (i.e., B molecules are desorbed). Thus $p_i = q_i \in \{1,0\}$, and the model reduces to the two-state lattice-gas model for single-component adsorption of A. 
The corresponding effective $S=1/2$ Ising model has 
\begin{equation}
\begin{split}
\sigma_i &= 2q_i -1 = 2p_i -1 \\
\hat{J}&= \cfrac{1}{4} (J+K+2L) = \cfrac{1}{4} \phi_{AA} \\
\hat{H}& = \cfrac{1}{2}(H-D) +  (J+K+2L) = \cfrac{1}{2}(\mu_A + 2 \phi_{AA}) \;.
\end{split}
\end{equation}

Analogously, as $H \to - \infty$ and $D\to +\infty $  $ (\mu_A \to -\infty)$, 
then $p_i = -q_i \in \{-1,0 \}$. Molecules of type A become energetically 
unfavorable and are desorbed. 
The model reduces to the two-state lattice-gas model for single-component adsorption of B. 
The corresponding effective $S=1/2$ Ising model has
\begin{equation}
\begin{split}
\sigma_i &= 2q_i -1 = -2p_i -1\\
\hat{J}&= \cfrac{1}{4} (J+K-2L) = \cfrac{1}{4} \phi_{BB} \\
\hat{H}& = \cfrac{1}{2}(-H-D) +  (J+K-2L) = \cfrac{1}{2}(\mu_B + 2 \phi_{BB}) \;.
\end{split}
\end{equation}

We now have the information necessary to construct the complete set of topologically 
different ground-state diagrams and determine their respective stability conditions.  

\subsection{Topologically different ground-state diagrams}
\label{topol}

Starting from one of the three strong-field limits, $H \to \pm \infty$ and $D \to +\infty$, 
and proceeding toward one of the other two, one can determine which, if any, 
of the ordered states first becomes lower in energy than the uniform states. 
The values of the fields $H$ and $D$ that mark the transitions between different ground states 
depend on the values of the interaction parameters $J,K,$ and $L$. 
Different values of  these parameters therefore lead to topologically different 
ground-state diagrams in the $\{ H,D \}$ plane.

In the rest of this paper we will for convenience 
sometimes use the normalized interactions and fields:  
$j = J/|J|$, $k = K/|J|$,  $\ell = L/|J|$, $h=H/|J|$, and  $d = D/|J|$. 
We now proceed to show that, in terms of $d$ and $h$, there are fifteen 
topologically different ground-state diagrams.  

Our first classification is in terms of the ground states for the FM $(j=+1)$ and 
AFM $(j = -1)$ cases in the $d \to - \infty$ limit. 
In the FM case, there is a direct transition between the uniform $(1 \times 1)_1^1$ and 
$(1 \times 1)_{-1}^1$ ground states at $h=-4 \ell$, 
while in the AFM case these uniform phases are separated 
by a region of the checkerboard $(\sqrt{2}\times\sqrt{2})_{0}^{1}$ (mixed A and B) phase 
in the range $ 4j -4 \ell < h < -4 \ell - 4j$. 

\begin{figure}
\begin{flushleft}
\hspace*{-20mm}
\includegraphics[scale=0.6,width=20cm,height=15cm]{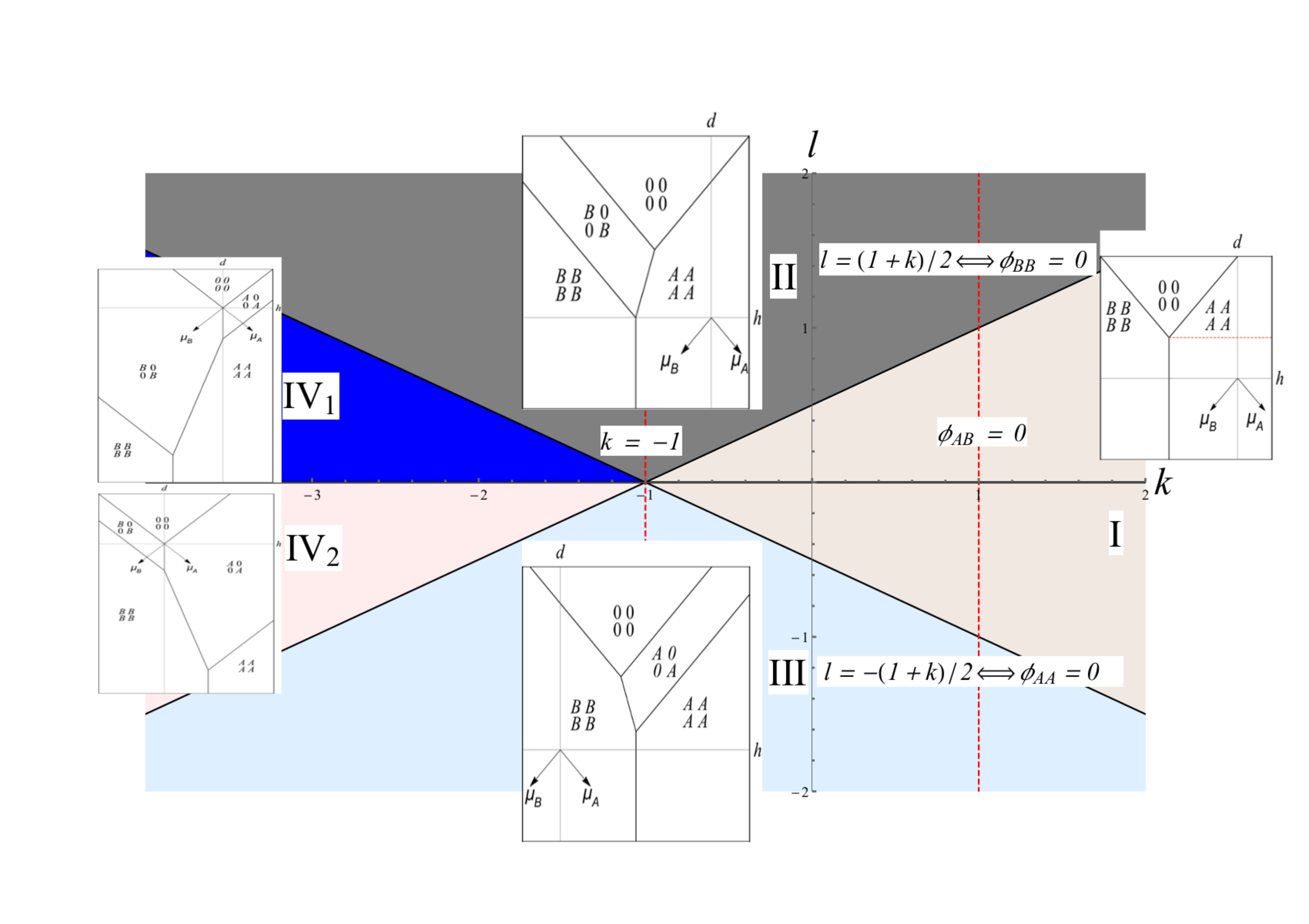}
\end{flushleft}
\vspace*{-15mm}
\caption{The phase diagram in the \{$k, \ell$\} parameter plane for the FM case. 
The different colors denote regions corresponding to topologically different 
ground-state diagrams in 
the \{$ h,d$\} plane. The ground-state diagrams are inset in each region. 
The line $k=1$ ($\phi_{AB} = 0$) corresponds to the lower left half-plane in 
Fig.~3 of Ref.\ \cite{FEFE18}.
}
\label{fig:fmgs}
\end{figure}

\begin{figure}
\begin{flushleft}
\hspace*{-20mm}
\includegraphics[scale=1,width=20cm,height=20cm]{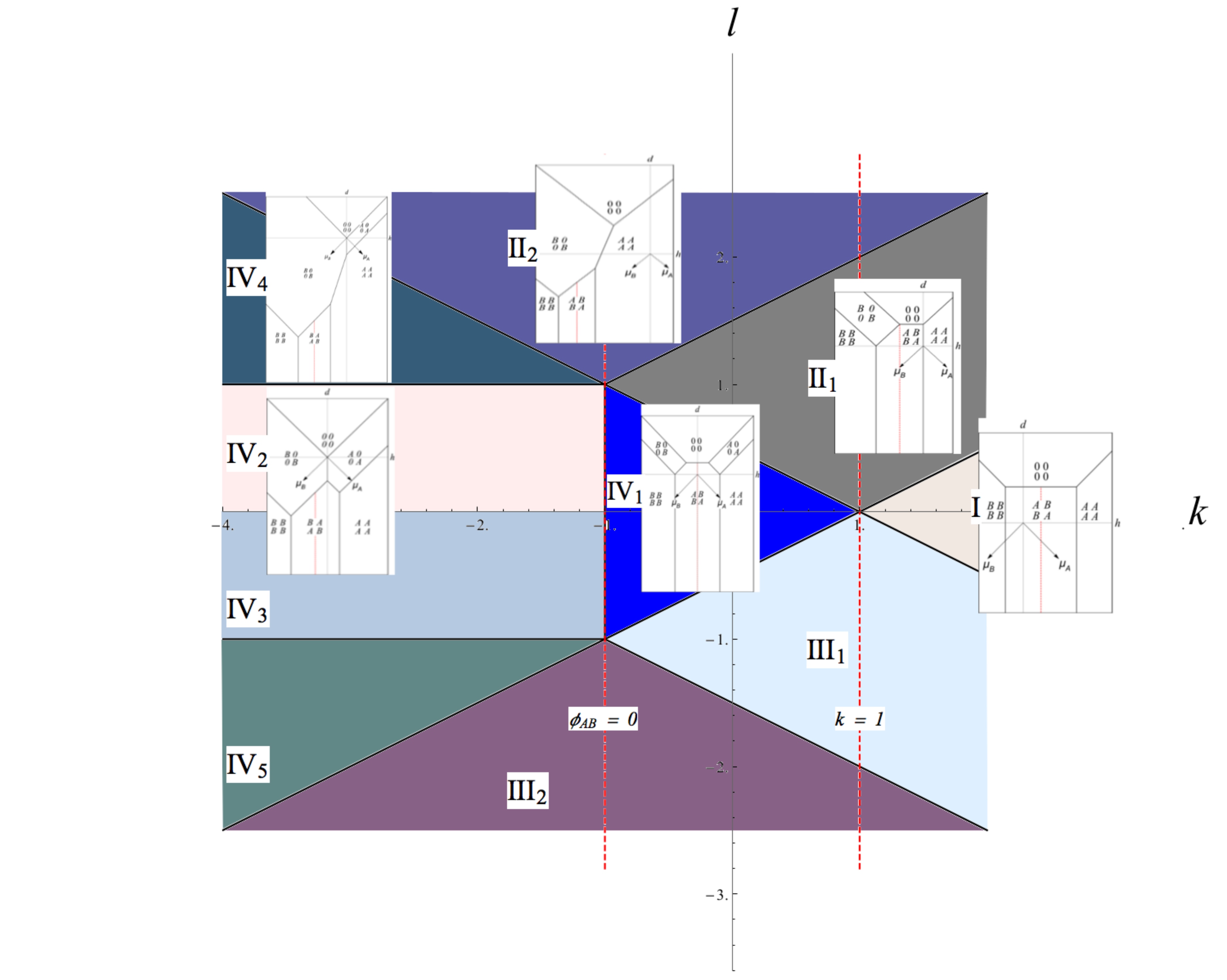}
\end{flushleft}
\vspace*{-15mm}
\caption{The phase diagram in the \{$k, \ell$\} parameter plane for the AFM case. 
The different colors denote regions with topologically different ground-state diagrams 
in the \{$ d,h$\} plane. The ground-state diagrams are inset in each region except 
$III_1$, $III_2$,  $IV_3$, and $IV_5$ since they are symmetric to $II_1$, $II_2$, $IV_2$, 
and $IV_4$ respectively under the transformation $h \to -h , l \to -l \iff A \leftrightarrow B $. 
The line $k=-1$ ($\phi_{AB} = 0$) corresponds to the upper right half-plane in 
Fig.~3 of Ref.\ \cite{FEFE18}.
}
\label{fig:afmgs}
\end{figure}

In both the FM and AFM cases, the $\{k,\ell\}$ plane is divided into four 
main regions, denoted I-IV. 
The inequalities that define these four regions are shown in Table \ref{table:inequal} 
in both Ising and LG language. After this division of the $\{k,\ell\}$ plane into four main regions, 
the remaining subdivisions are made according 
to the different topologies the ground-state diagrams can have in the $\{h,d\}$ plane. In Fig.\ \ref{fig:fmgs} we show the five regions in the $\{ k,\ell \}$ plane that correspond to topologically distinct ground states in the FM case. Likewise, in Fig.\ \ref{fig:afmgs} 
we show the ten regions in the $\{k,\ell \}$ plane that correspond to topologically distinct 
ground-state idagrams in the AFM case.
In both cases, the lines that separate different regions correspond to degenerate 
ground-state diagrams, intermediate between the adjoining topologies. 
In these figures, the $k$ axis corresponds to the BEG model, and the origin to the BC model. 

\begin{center}  
\begin{table}
\caption{Main regions in the \{$k,\ell$\} plane.}
\label{table:inequal}
  \begin{tabular}{ l  c  c }
    \hline
    \hline
    Region & Ising condition & Lattice-gas condition \\ \hline
     I & $  -1-k<2 \ell < 1+k $ & $\phi_{AA} >0, \phi_{BB} >0$ \\ \hline
     II & $  2\ell > -1-k \text{ and  } 2 \ell > 1+k $ & $\phi_{AA} >0, \phi_{BB} <0$ \\ \hline
     III & $  2\ell < -1-k \text{ and  } 2 \ell < 1+k $ & $\phi_{AA} <0, \phi_{BB} >0$ \\ \hline
     IV & $   1+k <2\ell < -1-k $ & $\phi_{AA} <0, \phi_{BB} <0$\\
       \hline
    \hline
  \end{tabular}
\end{table}
\end{center}

%\begin{figure}
%\begin{center}
%]\includegraphics[scale=0.6]{figures/}
%\end{center}
%\\caption{$D\to -\infty $}
%\label{fig:Dinf}
%\end{figure}

The four main regions in the $\{ k,\ell \} $ plane for both 
the FM or AFM cases correspond to the asymptotic behaviors for large $-\mu_A$ or $-\mu_B$. 
In both cases, this causes the Hamiltonian to reduce to an effective $S=1/2$ Ising model. 
For large $-\mu_A$ the ground-state diagram is symmetric about 
$\mu_B = -2(J+K-2L) = -2 \phi_{BB}$. 
For large $-\mu_B$ the ground-state diagram is symmetric about the line 
$\mu_A = -2(J+K+2L) = -2 \phi_{AA}$. 
Each of these limits 
may be FM or AFM in the $D \to - \infty$ limit, depending on whether $J$ is positive or negative. 
This leads to four different ways to combine the asymptotic behaviors for $-\mu_A$ and $-\mu_B$. 

In Fig.\ \ref{fig:fmgs} we show the five regions in the $\{ k,\ell \}$ plane for the FM case, with 
examples of the corresponding ground-state diagrams inserted. 
The line $k=1$ ($\phi_{AB} = 0$), which cuts through sections I, II, and III, corresponds to the lower left half-plane in Fig.~3 of Ref.\ \cite{FEFE18}. 
In the FM case ($j=+1$), this restriction excludes the possibility of 
both $\phi_{AA}$ and $\phi_{BB}$ being 
negative, which corresponds to main section IV in this figure. 

In Fig.\ \ref{fig:afmgs} we show the ten regions in the $\{ k,\ell \}$ plane for the AFM case, with 
examples of the corresponding ground-state diagrams inserted. 
The ground-state diagram for the AFM BC model corresponds to the origin. It was previously 
presented in Ref.\ \cite{KIME91} in the context of a study of the tricritical properties of this model 
at nonzero temperatures. 
The line $k=-1$ ($\phi_{AB} = 0$), which cuts through sections II$_2$ and III$_2$ 
and corresponds to the border between section IV$_1$ and sections IV$_2$ and IV$_3$,
corresponds to the upper right half-plane in Fig.~3 of Ref.\ \cite{FEFE18}. 
It does not include the topologically different ground-state diagrams in sections I, II$_1$, 
III$_1$, IV$_4$, and IV$_5$. 

In both Figs.\ \ref{fig:fmgs} and \ref{fig:afmgs}, the equations that mark the transition lines 
between pairs of ground states in the $\{ h,d \}$ plane are omitted for readability. 
They are calculated by pairwise equating the energies per site that are given in 
Table \ref{table:gs}, and they are shown in Table \ref{table:pt}.

\begin{turnpage}
\begin{table}
\caption{Equations for direct transition lines between pairs of ground states. 
The table has the form of a symmetric matrix. 
The ground state $(\sqrt{2}\times\sqrt{2})_{0}^1 $ exists only in the AFM case ($J < 0$).}
%\caption{Possible Transitions}
\label{table:pt}
\resizebox{\linewidth}{!}{
  \begin{tabular}{| l |  c | c | c | c | c | c |}
    \hline
    \hline
    State & $ (1\times1)_1^1 $ & $ (1\times1)_{-1}^1 $ & $(1\times1)_{0}^0 $ & $(\sqrt{2}\times\sqrt{2})_{0}^1 $ & $(\sqrt{2}\times\sqrt{2})_{1/2}^{1/2} $  & $(\sqrt{2}\times\sqrt{2})_{-1/2}^{1/2}$ \\ \hline
    $(1\times1)_1^1 $&  & $H = -4 L$ &  $D = H + 2(J + K+ 2 L) $ & $H= -4L + 4|J|$ & $D=H+4(J+K+2L)$ & $D=3H+4(J+K+2L)$ \\ \hline
    $(1\times1)_{-1}^1 $ & $H = -4 L$ &  & $D = -H + 2(J + K- 2L) $ &  $H= -4L - 4|J|$ & $D=-3H+4(J+K-2L)$ & $D=-H+4(J+K-2L)$ \\ \hline
    $(1\times1)_{0}^0 $ &  $D = H + 2(J + K+ 2 L) $ &  $D = -H + 2(J + K- 2L) $ &  & $D= 2(|J| +K)$ & $D=H$ &  $D= -H$\\ \hline
     $(\sqrt{2}\times\sqrt{2})_{0}^1 $ & $H= -4L + 4|J| $ &  $H= -4L - 4|J|$ & $D= 2(|J| +K)$ & & $D = -H + 4(|J| + K)$ & $D = H + 4(|J| + K)$\\ \hline     
      $(\sqrt{2}\times\sqrt{2})_{1/2}^{1/2} $ & $D=H+4(J+K+2L)$  & $D=-3H+4(J+K-2L)$  &  $D= H $ &  $D = -H + 4(|J| + K)$ &  &$H=0$\\ \hline
      $(\sqrt{2}\times\sqrt{2})_{-1/2}^{1/2} $ & $D=H+4(J+K+2L)$ & $D=-H+4(J+K-2L)$ & $ D= -H $ & $D = H + 4(|J| + K)$ &$H=0$  & \\
    \hline
    \hline
  \end{tabular}
  }
\end{table}
\end{turnpage}

\section{Summary and Conclusions}
\label{sec:CONC}

The ground-state diagram for a statistical-mechanical model provides a solid foundation for 
studies of equilibrium and nonequilibrium phenomena at nonzero temperature. 
In this paper we therefore present a complete catalog of the fifteen topologically different 
ground-state diagrams for the most general 
three-state lattice-gas 
or equivalently $S=1$ Ising 
model with 
only nearest-neighbor interactions on a square lattice. 
This model is often 
used to study phase transitions in 
two-component adsorption at solid-gas and solid-liquid interfaces, as well as in 
magnetic and dielectric spin systems, and it 
provides a rich laboratory for studying a number of critical and multicritical phenomena within the 
framework of one single model.  
The square lattice has important physical realizations as the (100) planes of face-centered and 
body-centered cubic crystals, and it is also often used as a simple basis for theoretical studies. 

The model is defined 
in a five-dimensional parameter space consisting of three interaction constants and 
two external chemical potentials or fields. Six topologically different ground-state diagrams for this 
model in a subspace with only two independent interaction constants were recently published 
\cite{FEFE18}, but we are not aware of previous publication of a complete catalog for the 
full, five-dimensional parameter space. 
We thus feel that our results fill a void, and we hope that they will be useful for future 
research in physical chemistry and chemical physics at interfaces.

\section*{Conflicts of interest}
There are no conflicts to declare.

\section*{Acknowledgments}
This material is based upon work supported by the U.S. Department of Energy Office of Science, 
Office of Nuclear Physics under Award Number DE-FG02-92ER40750. 
P.A.R.\ acknowledges partial support by U.S.\ National Science Foundation Grant No. DMR-1104829. 
Work at the University of Oslo was partly supported by the Research Council of Norway 
through the Center of Excellence funding scheme, Project No. 262644.

%\clearpage
%\bibliography{GSpaper}
%\bibliography{elchem}
%\bibliography{metastab}
%\bibliographystyle{prsty}
%\bibliographystyle{unsrt}
%\bibliographystyle{apsrev}
%\clearpage
%\clearpage

\end{document}